\documentclass[%
 preprint,
 amsmath,amssymb,
 aps,
 pre,
 showkeys
]{revtex4-2}
\usepackage{amsmath}
\usepackage[utf8]{inputenc}
\usepackage{graphicx}
\usepackage{dcolumn}
\usepackage{bm}
\usepackage{hyperref}
\usepackage{xcolor}
\usepackage{url}

\begin{document}

\title{Wall tension of a Bose-Einstein condensate: Effects of quantum
  fluctuations and finite-range interactions}

\author{Pham Duy Thanh}
\author{Nguyen Van Thu}
\email{nvthu@live.com}
\affiliation{Department of Physics, Hanoi Pedagogical University 2,
  Hanoi, Vietnam}

\date{\today}

\begin{abstract}
We investigate the effects of quantum fluctuations (QFs) and finite-range interatomic interactions on the ground state and wall tension of a Bose-Einstein condensate (BEC) at zero temperature by means of modified Gross-Pitaevskii equations. The QFs are shown to stiffen the condensate near the hard
wall, narrowing the density profile and raising the wall tension above its mean-field value by an amount proportional to square-root of the gas parameter. Finite-range corrections always reduces the wall tension relative to the QFs correction value regardless of the effective range.  We find the ordering $\gamma_0 < \gamma_{\rm II} < \gamma_{\rm I}$ in the dilute regime and identify a crossover condition under which the two beyond-mean-field contributions mutually cancel.  These results provide an experimentally accessible way for probing beyond-LHY physics through precision measurements of the wall tension in ultracold Bose gases.
\end{abstract}


\maketitle


\section{Introduction}
\label{sec1}

When cooled below a critical temperature, a dilute gas of bosonic atoms undergoes a phase transition from the normal phase to a Bose-Einstein condensate (BEC)~\cite{Pitaevskij,Pethick}.  At zero temperature, all atoms theoretically reside in the single-particle ground state, and the
many-body system is described by a macroscopic wave function.  Within the standard Gross-Pitaevskii (GP) theory, this wave function satisfies a nonlinear Schr\"{o}dinger equation ~\cite{Gross,Pitaevskii}.  The GP equation has proven remarkably successful in describing both the static and dynamic properties of single-component as well as multicomponent BECs. However,
it is fundamentally a mean-field theory in which quantum fluctuations (QFs) are entirely neglected.

Even at zero temperature, a finite fraction of atoms is promoted from the condensate to excited states as a direct consequence of QFs.  Within the mean-field approximation, the GP theory ignores this quantum depletion, whereas diffusion Monte Carlo calculations have demonstrated that QFs can contribute at the level of $20\%$ to the wall tension~\cite{Blume2001}.  Beyond the GP framework, the Cornwall-Jackiw-Tomboulis (CJT) effective-action formalism~\cite{CJT} provides a systematic way to incorporate QFs into the description of a BEC at zero temperature~\cite{Thu2022}. The CJT approach is particularly well suited to homogeneous systems, where the condensate wave function is spatially uniform.  Nevertheless, both the GP theory and the CJT effective-action approach treat the interatomic
interaction as a contact potential whose coupling constant is evaluated within the first Born approximation~\cite{Pethick}, thereby neglecting the finite spatial extent of the interaction potential.

To account simultaneously for QFs and finite-range effects in the condensate wave function, Fu \textit{et al.}~\cite{fu2003beyond} proposed a modified Hamiltonian that goes beyond the Fermi
pseudopotential.  The principal aims of the present work are to extend this approach in order to study the influence of QFs and finite-range interactions on both the condensate wave function and the wall tension of a BEC confined by a hard wall at zero temperature.

The remainder of this paper is organized as follows.  In Sec.~\ref{sec2} we derive the condensate wave function at successive levels of approximation: GP, MGPI (QFs included), and MGPII (QFs and finite-range included).  Section~\ref{sec3} presents the calculation of the wall tension at each level and discusses the physical mechanisms underlying the beyond-mean-field corrections.  Conclusions are given in Sec.~\ref{sec4}.

\section{Wave function of condensate}
\label{sec2}

In this section we derive the condensate wave function incorporating, successively, the effects of the QFs and finite-range interactions. We consider a system of $N$ interacting bosons enclosed in a volume $V$ at zero temperature with no external potential.  The system is assumed to be translationally invariant in the $(y,z)$-plane, with inhomogeneity only along the $x$-axis. In the second-quantization
formulation, the Hamiltonian is
\begin{eqnarray}
\hat{H} &=& \int dx\,\hat{\Psi}^{\dagger}(x)
  \left[-\frac{\hbar^{2}}{2m}\frac{d^{2}}{dx^{2}}\right]\hat{\Psi}(x)
  \nonumber\\
&&{}+\frac{1}{2}\int dx_{1}\int dx_{2}\,
  \hat{\Psi}^{\dagger}(x_{1})\hat{\Psi}^{\dagger}(x_{2})
  U(x_{1}-x_{2})\hat{\Psi}(x_{2})\hat{\Psi}(x_{1}),
\label{Hamilton}
\end{eqnarray}
where $\hbar$ is the reduced Planck constant, $m$ is the atomic mass, and $\hat{\Psi}(x)$, $\hat{\Psi}^{\dagger}(x)$ are the bosonic field annihilation and creation operators, respectively. For a short-range two-body interaction, the potential is modeled by the contact form
\begin{eqnarray}
U(x_{1}-x_{2}) = g\,\delta(x_{1}-x_{2}),
\label{potential}
\end{eqnarray}
where the coupling constant $g$, evaluated in the first Born approximation, is related to the $s$-wave scattering length $a_{s}$ by
\begin{eqnarray}
g = \frac{4\pi\hbar^{2}a_{s}}{m}.
\label{g}
\end{eqnarray}

Within mean-field theory, the state of the system is described by the order parameter $\psi_{0}(x)\equiv\langle\hat{\Psi}(x)\rangle$, which satisfies the GP equation
\begin{eqnarray}
-\frac{\hbar^{2}}{2m}\frac{d^{2}\psi_{0}}{dx^{2}}
  - \mu_{0}\psi_{0} + g\psi_{0}^{3} = 0,
\label{GP0}
\end{eqnarray}
with the mean-field chemical potential $\mu_{0}=gn_{0}\equiv gN/V$. Introducing the healing length $\xi$, the dimensionless coordinate $\tilde{x}$, and the reduced wave function $\tilde{\psi}_{0}$ via
\begin{eqnarray}
\xi  &=& \frac{\hbar}{\sqrt{2mgn}},\nonumber\\
\tilde{x} &=& \frac{x}{\xi},\qquad
\tilde{\psi}_{0} = \frac{\psi_{0}}{\sqrt{n}},
\label{dimension}
\end{eqnarray}
the GP equation~(\ref{GP0}) takes the dimensionless form
\begin{eqnarray}
-\frac{d^{2}\tilde{\psi}_{0}}{d\tilde{x}^{2}}
  - \tilde{\psi}_{0} + \tilde{\psi}_{0}^{3} = 0.
\label{GP01}
\end{eqnarray}
Assuming that the system is rendered inhomogeneous by a hard-wall located at $\tilde{x}=0$, the boundary condition is
\begin{eqnarray}
\tilde\psi_0(0) = 0,~\tilde\psi_0(\infty)=1.
\label{BC}
\end{eqnarray}
Equations~(\ref{GP01}) and~(\ref{BC}) admit the well-known exact solution
\begin{eqnarray}
\tilde{\psi}_{0}(\tilde{x}) =
  \tanh\!\left(\frac{\tilde{x}}{\sqrt{2}}\right).
\label{solution0}
\end{eqnarray}

\subsection{Quantum-fluctuation correction: MGPI equation}
\label{sec2a}

To incorporate quantum fluctuations we employ the energy functional proposed in Ref.~\cite{Fabrocini2001}, which augments the GP functional with the  Lee, Huang and Yang (LHY) energy density~\cite{fu2003beyond}
\begin{eqnarray}
E[\psi]_{\rm MGPI}
  &=& E_{\rm GP}
    + \frac{g}{2}\int d\vec{r}\,
      \tilde{\mathcal{E}}_{\rm LHY}\,\psi^{4}|\psi|
  \nonumber\\
  &=& \int d\vec{r}\left[
    \frac{\hbar^{2}}{2m}|\nabla\psi|^{2}
    + \frac{g}{2}|\psi|^{4}
    \!\left(1+\tilde{\mathcal{E}}_{\rm LHY}|\psi|\right)
    \right],
\label{EMGPI}
\end{eqnarray}
where $\tilde{\mathcal{E}}_{\rm LHY}=\tfrac{128}{15\sqrt{\pi}}\alpha_{s}^{1/2}$ is the LHY coefficient first derived in Refs.~\cite{lee1957eigenvalues,lee1957many} and later recovered within the CJT framework~\cite{Thu2022}.  Here $\alpha_{s}=na_{s}^{3}\ll 1$ is the gas parameter characterizing a dilute BEC~\cite{Pethick}. Minimizing the functional~(\ref{EMGPI}) with respect to $\psi$ yields
the first modified GP equation (MGPI),
\begin{eqnarray}
-\frac{\hbar^{2}}{2m}\frac{d^{2}\psi_{\rm I}}{dx^{2}}
  - \mu_{\rm I}\psi_{\rm I}
  + g\psi_{\rm I}^{3}
  + g_{\rm I}\psi_{\rm I}^{4} = 0,
\label{GPI}
\end{eqnarray}
with
\begin{eqnarray}
g_{\rm I} = \frac{32}{3\sqrt{\pi}}\,g\,a_{s}^{3/2}.
\label{gI}
\end{eqnarray}
Multiplying Eq.~(\ref{GP0}) by $\psi_{0}^{*}$, multiplying Eq.~(\ref{GPI}) by $\psi_{\rm I}^{*}$,
integrating over the volume, and subtracting gives 
\begin{eqnarray}
\mu_{\rm I} - \mu_{0}
  = \frac{g_{\rm I}}{N}\int|\psi_{\rm I}|^{5}\,dx.
\label{hieumu}
\end{eqnarray}
To leading order in $\alpha_{s}^{1/2}$ this yields the shifted chemical potential
\begin{eqnarray}
\mu_{\rm I} = \mu_{0}(1+\delta\tilde{\mu}),\qquad
\delta\tilde{\mu} = \frac{32\alpha_{s}^{1/2}}{3\sqrt{\pi}}.
\label{muI}
\end{eqnarray}

Far from the hard wall, where the condensate is uniform, the energy density extracted from Eq.~(\ref{EMGPI}) is
\begin{eqnarray}
\mathcal{E} = \frac{1}{2}gn_{c}^{2} + \mathcal{E}_{\rm LHY},
\label{calE}
\end{eqnarray}
where $n_{c}$ is the condensate density and the LHY contribution reads
\begin{equation}
\mathcal{E}_{\rm LHY}
  = \frac{gn_{c}^{2}}{2}\,\tilde{\cal E}_{\rm LHY}.
\label{LHY}
\end{equation}
Exploiting the $U(1)$ Ward identity~\cite{Floerchinger2008}, the non-condensed (quantum-depleted) fraction is
\begin{eqnarray}
n_{\rm ex} \equiv n - n_{c}
  = \frac{1}{4}\frac{\partial\mathcal{E}_{\rm LHY}}{\partial(gn_{c})},
\label{ward}
\end{eqnarray}
which gives the standard LHY depletion result,
\begin{eqnarray}
\frac{n_{\rm ex}}{n} = \frac{8}{3\sqrt{\pi}}\,\alpha_{s}^{1/2}.
\label{noncondensed}
\end{eqnarray}
The results for chemical potential (\ref{muI}) and non-condensed fraction (\ref{noncondensed}) are in agreement with those obtained via the CJT effective-action approach~\cite{Thu2022}, the variational
method~\cite{Stringari2018}, and the pseudopotential formalism~\cite{lee1957eigenvalues}.

To study the local condensate density along the $x$-axis, we use the rescalings~(\ref{dimension}) with $n$ replaced by $n_{c}$, so that the MGPI equation~(\ref{GPI}) is rewritten in dimensionless form
\begin{eqnarray}
-\frac{d^{2}\tilde{\psi}_{\rm I}}{d\tilde{x}^{2}}
  - (1+\delta\tilde{\mu})\tilde{\psi}_{\rm I}
  + \tilde{\psi}_{\rm I}^{3}
  + \delta\tilde{\mu}\,\tilde{\psi}_{\rm I}^{4} = 0.
\label{GPI1}
\end{eqnarray}
The boundary condition now becomes
\begin{eqnarray}
\tilde\psi_{\rm I}(0)=0,~\tilde\psi_{\rm I}(\infty)=1.\label{BCI}
\end{eqnarray}

For a dilute gas with $\alpha_{s}\ll 1$, the solution may be expanded perturbatively in the gas parameter.  To first order one writes
\begin{eqnarray}
\tilde{\psi}_{\rm I}
  \approx \tilde{\psi}_{0}
    + \frac{32\alpha_{s}^{1/2}}{3\sqrt{\pi}}\tilde{\psi}_{1}
    + \mathcal{O}(\alpha_{s}).
\label{expand1}
\end{eqnarray}
Substituting~(\ref{expand1}) into~(\ref{GPI1}) and retaining only terms linear in $\alpha_{s}^{1/2}$ yields the differential equation for the first-order correction $\tilde{\psi}_{1}$,
\begin{eqnarray}
\left(-\frac{d^{2}}{d\tilde{x}^{2}} - 1
  + 3\tilde{\psi}_{0}^{2}\right)\tilde{\psi}_{1}
  = \tilde{\psi}_{0} - \tilde{\psi}_{0}^{4}.
\label{y1}
\end{eqnarray}
The solution of~Eq. (\ref{y1}) is
\begin{eqnarray}
\frac{32\alpha_{s}^{1/2}}{3\sqrt{\pi}}\tilde{\psi}_{1}
  = \frac{8\alpha_{s}^{1/2}}{3\sqrt{\pi}}
    \frac{e^{-\sqrt{2}\tilde{x}}}
         {15\!\left(1+e^{\sqrt{2}\tilde{x}}\right)^{2}}
    F(\tilde{x}),
\label{y11}
\end{eqnarray}
with
\begin{eqnarray*}
F(\tilde{x}) &=& 1 - 12e^{\sqrt{2}\tilde{x}}
  + e^{2\sqrt{2}\tilde{x}}\!\left[
    -2\ln\!\left(e^{\sqrt{2}\tilde{x}}\right)
    +32\ln\!\left(e^{\sqrt{2}\tilde{x}}+1\right)
    +11 - 32\ln 2
    \right].
\end{eqnarray*}

\begin{figure}[ht]
\centering
\includegraphics[width=0.6\textwidth]{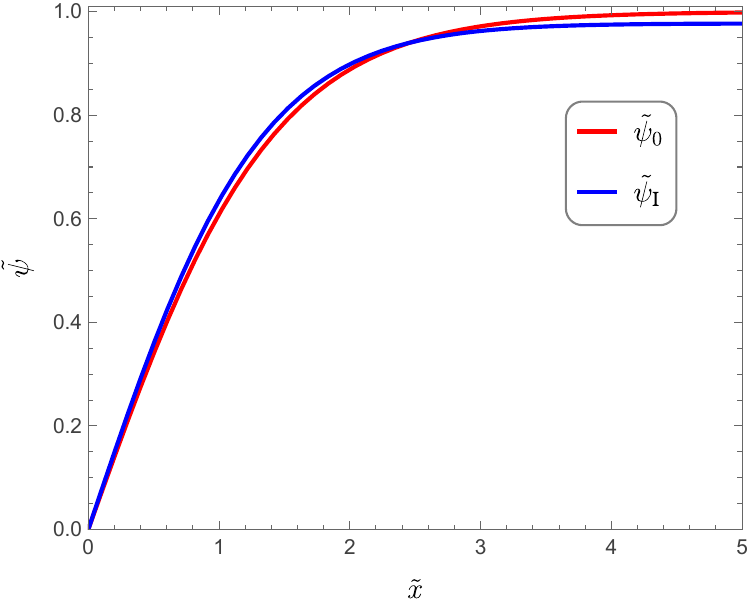}
\caption{Condensate wave functions as functions of the reduced
  coordinate $\tilde{x}$ at $\alpha_{s}=10^{-3}$.  The blue and red
  curves correspond to the GP~(\ref{solution0}) and
  MGPI~(\ref{expand1}) solutions, respectively.}
\label{f:fig1}
\end{figure}

The GP and MGPI wave functions are compared in Fig.~\ref{f:fig1} at $\alpha_{s}=10^{-3}$.  The QFs are seen to increase the condensate density in the vicinity of the hard wall.  Within the GP theory, the density profile recovers monotonically from zero at the wall to unity in the bulk, with a characteristic length scale set by the healing length $\xi$.  When the LHY correction is included, the
MGPI solution satisfies $\tilde{\psi}_{\rm I}>\tilde{\psi}_{0}$ near the hard-wall subject to the same boundary conditions $\tilde{\psi}_{\rm I}(0)=0$ and $\tilde{\psi}_{\rm I}(+\infty)=1$. The physical mechanism is transparent: the LHY term $g_{\rm I}\psi_{\rm I}^{4}$ in Eq.~(\ref{GPI}) introduces a
density-dependent repulsion that stiffens the condensate and raises the chemical potential from $\mu_{0}$ to the value given by Eq.~(\ref{muI}).  A larger chemical potential entails a higher
energetic penalty for maintaining a region of suppressed density, so the condensate recovers more rapidly from the low-density region near the hard-wall.  Consequently, the density deficit $1-\tilde{\psi}_{\rm I}^{2}$ is smaller than its GP counterpart at every position, and the
density notch is simultaneously shallower and narrower.  This behavior does not conflict with quantum depletion: the depleted fraction is already encoded in the renormalized chemical potential and coupling
constant entering the MGPI equation, and does not alter the asymptotic boundary conditions on the wave function.

An alternative approximate solution of Eq.~(\ref{GPI1}) can be obtained by using linear approximation ~\cite{Ao1998},
\begin{eqnarray}
\tilde{\psi}_{\rm I}
  \approx \tanh\!\left[
    \frac{\tilde{x}}{\sqrt{2}}
    \left(1+\frac{3\delta\tilde{\mu}}{4}\right)
    \right].
\label{solution2}
\end{eqnarray}
Restoring physical units, the healing length modified by the QFs is
\begin{eqnarray}
\xi_{\rm I} =
  \frac{\xi}{1+\tfrac{3}{4}\delta\tilde{\mu}},
\label{xiI}
\end{eqnarray}
confirming that QFs shorten the healing length relative to its GP value, consistent with the narrowing of the wave-function profile seen in Fig.~\ref{f:fig1}.

\subsection{Finite-range correction: MGPII equation}
\label{sec2b}

In the standard GP theory, both QFs and the finite spatial extent of the interatomic potential are neglected. The MGPI equation accounts for the former; we now additionally include the
finite-range correction.  Following Ref.~\cite{fu2003beyond}, the resulting equation (MGPII) is
\begin{eqnarray}
-\frac{\hbar^{2}}{2m}\frac{d^{2}\psi_{\rm II}}{dx^{2}}
  - \mu_{\rm II}\psi_{\rm II}
  + g\psi_{\rm II}^{3}
  + g_{\rm I}\psi_{\rm II}^{4}
  + \frac{g_{2}}{2}\,\psi_{\rm II}
    \frac{d^{2}(\psi_{\rm II}^{2})}{dx^{2}} = 0,
\label{GPII}
\end{eqnarray}
where
\begin{eqnarray}
g_{2} = \frac{4\pi\hbar^{2}a_{s}^{2}(a_{s}-r_{e}/2)}{m}.
\label{g2}
\end{eqnarray}
The quantity $r_{e}$ is the effective range of the two-body interaction potential, which makes $g_2$ may be positive or negative depending on the microscopic interaction model. Because the finite-range correction to the chemical potential is proportional to $\alpha_{s}^{3/2}$ and is
therefore negligible in the dilute limit~\cite{zhang2024cornwall}, we set $\mu_{\rm II}=\mu_{\rm I}$.  Using the rescalings~(\ref{dimension}), Eq.~(\ref{GPII}) reduces to the dimensionless form
\begin{eqnarray}
(b-1)\frac{d^{2}\tilde{\psi}_{\rm II}}{d\tilde{x}^{2}}
  + b\!\left(\frac{d\tilde{\psi}_{\rm II}}{d\tilde{x}}\right)^{2}
  - (1+\delta\tilde{\mu})\tilde{\psi}_{\rm II}
  + \tilde{\psi}_{\rm II}^{3}
  + \delta\tilde{\mu}\,\tilde{\psi}_{\rm II}^{4} = 0,
\label{GPII1}
\end{eqnarray}
with
\begin{eqnarray}
b = \frac{a_{s}^{2}}{\xi^{2}}\!\left(1-\frac{r_{e}}{2a_{s}}\right).
\label{b}
\end{eqnarray}
For most models of two-body interactions $b\ll 1$, and Eq.~(\ref{GPII1}) admits the perturbative solution
\begin{eqnarray}
\tilde{\psi}_{\rm II}
  = \tilde{\psi}_{\rm I}
  - \frac{b}{5}
    \frac{\ln\!\left(\cosh\frac{\tilde{x}}{\sqrt{2}}\right)}
         {\cosh^{2}\!\frac{\tilde{x}}{\sqrt{2}}}.
\label{psiII}
\end{eqnarray}
\begin{figure}[ht]
\centering
\includegraphics[width=0.6\textwidth]{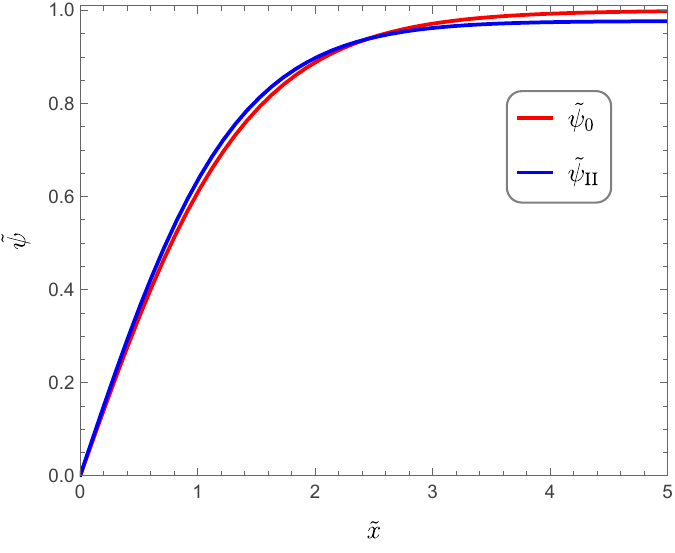}
\caption{Condensate wave functions as functions of the reduced coordinate $\tilde{x}$ at $\alpha_{s}=10^{-3}$ and $r_e=2a_s/3$.  The red and blue curves correspond to the GP~(\ref{solution0}) and MGPII~(\ref{psiII}) solutions, respectively.}
\label{f:fig2}
\end{figure}
The sign of the finite-range parameter $b$ is governed by the ratio $r_{e}/2a_{s}$. When $r_{e}<2a_{s}$, one has $b>0$, and the finite-range correction lowers the order parameter relative to the MGPI prediction, leading to a density notch that is deeper and wider than that obtained with LHY corrections alone. Conversely, for $r_{e}>2a_{s}$, the parameter $b$ becomes negative, and the finite-range contribution enhances the LHY-induced stiffening, resulting in a soliton that is shallower and narrower than the MGPI solution. This sign change admits a clear physical interpretation: the effective range encodes the momentum dependence of the two-body scattering amplitude, and when it is large compared to the scattering length, it suppresses the effective repulsion at momentum scales set by the healing length. This effect is particularly pronounced near Feshbach resonances, where $a_s$ can be tuned through zero and $r_e$ may become large and negative for narrow resonances~\cite{Inouye1998}, thereby rendering the regime $r_{e}>2a_{s}$ experimentally accessible and qualitatively modifying the soliton profile relative to both the GP and MGPI predictions.

Fig.~\ref{f:fig2} also shows the MGPII wave function at
$\alpha_{s}=5\times10^{-3}$ with $r_{e}=2a_{s}/3$, corresponding to
the hard-sphere potential. The difference between $\tilde\psi_{\rm II}$ and $\tilde\psi_{\rm I}$ in the vicinity of the hard wall is too small to be resolved in the same figure.

\section{Wall tension}
\label{sec3}

To compute the wall tension of the BEC bounded by a hard wall, we
start from the grand potential~\cite{Deng2016},
\begin{eqnarray}
\Omega = 2P\xi A\int_{0}^{\infty}d\tilde{x}\left[
  \left(\frac{\partial\tilde{\psi}_{i}}{\partial\tilde{x}}\right)^{2}
  + \tilde{V}_{i}^{2}\right],
\label{Omega}
\end{eqnarray}
where $A$ is the surface area, $P=gn^{2}/2$ is the bulk pressure,
and $\tilde{V}_{i}\equiv V_{i}/P$ ($i=0,\,{\rm I},\,{\rm II}$)
is the dimensionless effective potential.  In the GP theory, the dimensionless GP potential has the form
\begin{eqnarray}
\tilde V_0=-\tilde\psi_0^2+\frac{\tilde\psi_0^4}{2},\label{VGP}
\end{eqnarray}
which satisfies the "constant of motion" ~\cite{Indekeu2015}
\begin{eqnarray}
\left(\frac{\partial\tilde{\psi}_0}{\partial\tilde{x}}\right)^{2}
  - \tilde{V}_0 = \frac{1}{2}.
\label{const0}
\end{eqnarray}
The wall tension is defined as the excess grand-potential energy per
unit area,
\begin{eqnarray}
\gamma_0
  = \frac{\Omega + PV}{A}
  = 4P\xi\int_{0}^{\infty}d\tilde{x}
    \left(\frac{\partial\tilde{\psi}_0}{\partial\tilde{x}}\right)^{2}.
\label{gamma0}
\end{eqnarray}
Substituting the GP solution~(\ref{solution0}) into Eq.~(\ref{gamma0})
gives the well-known result for wall tension \cite{Barankov2002,Schaeybroeck2008,Indekeu2015,Deng2016}
\begin{eqnarray}
\gamma_{0} = \frac{4\sqrt{2}}{3}\,P\xi.
\label{gamma01}
\end{eqnarray}
This result, which reflects the purely mean-field balance between kinetic and interaction energy.

\subsection{MGPI result: LHY correction}

Within presence of the QFs, instead of (\ref{VGP}), the corresponding modified GP potential is
\begin{eqnarray}
\tilde V_{\rm I}=-(1+\delta\tilde\mu)\tilde\psi_{\rm I}^2+\frac{\tilde\psi_{\rm I}^4}{2}+\frac{\delta\tilde\mu}{5}\tilde\psi_{\rm I}^5.\label{VGPI}
\end{eqnarray}
Associating the boundary condition (\ref{BCI}), the "constant of motion" is read
\begin{eqnarray}
\left(\frac{\partial\tilde{\psi}_{\rm I}}{\partial\tilde{x}}\right)^{2}
  - \tilde{V}_{\rm I} = \frac{1}{2}+\tilde{\cal E}_{\rm LHY}.
\label{const1}
\end{eqnarray}
Plugging (\ref{const1}) into (\ref{Omega}) one can define the wall tension within presence of the QFs
\begin{eqnarray}
\gamma_{\rm I}
  = \frac{\Omega + P_{\rm I}V}{A}
  = 4P_{\rm I}\xi\int_{0}^{\infty}d\tilde{x}
    \left(\frac{\partial\tilde{\psi}_{\rm I}}{\partial\tilde{x}}\right)^{2},
\label{gamma1}
\end{eqnarray}
in which 
\begin{eqnarray}
P_{\rm I}=P_0\left(1+\frac{64}{15\sqrt{\pi}}\alpha_s^{1/2}\right).\label{PI}
\end{eqnarray}
The last term in bracket of right-hand side (\ref{PI}) corresponds to contribution of the QFs to the pressure and is called the dimensionless LHY pressure.

Substituting the MGPI wave function~(\ref{solution2}) into Eq.~(\ref{gamma1}) yields the beyond-mean-field wall tension at the
LHY level,
\begin{eqnarray}
\gamma_{\rm I}
  = \left(\frac{4\sqrt{2}}{3}
    + \frac{32\sqrt{2}}{3\sqrt{\pi}}\,\alpha_{s}^{1/2}\right)P_{\rm I}\xi.
\label{gammaI}
\end{eqnarray}
The LHY correction increases the wall tension above its mean-field value by an amount $\approx 8.511\,\alpha_{s}^{1/2}\,P\xi$. This behavior has a transparent physical origin: QFs stiffen the condensate against density modulations, thereby increasing the energetic cost of locally depleting the density below its bulk value. The resulting enhancement of stiffness also narrows the soliton profile, leading to steeper density gradients at the core and, consequently, larger kinetic-energy contributions; these effects act cooperatively. Since the LHY coefficient is positive and $\alpha_{s}\geq 0$ by definition, the inequality $\gamma_{\rm I}>\gamma_{0}$ holds throughout the dilute regime. This result is fully consistent with the general role of quantum fluctuations in weakly interacting Bose gases: beyond-mean-field corrections systematically increase the energetic cost of sustaining density interfaces, as exemplified by the enhanced surface tension of self-bound quantum droplets stabilized by LHY repulsion~\cite{Petrov2015}.

\subsection{MGPII result: finite-range correction}

The last term on the right-hand side of the MGPII equation~(\ref{GPII}) does not modify the constant of motion given in Eq.~(\ref{const1}), since the first derivative of the wave function vanishes in the bulk region. Substituting the MGPII wave function~(\ref{psiII}) into Eq.~(\ref{gamma1}) then yields the wall tension including both quantum fluctuations and finite-range effects,
\begin{eqnarray}
\gamma_{\rm II}
  = \gamma_{\rm I} - 0.0063\,b^{2}\,P\xi
  = \left(\frac{4\sqrt{2}}{3}
    + 8.511\,\alpha_{s}^{1/2}
    - 0.0063\,b^{2}\right)P\xi.
\label{gamma2}
\end{eqnarray}
The finite-range correction enters with a negative sign, reflecting a partial softening of the condensate response: the finite spatial extent of the interatomic potential counteracts the LHY stiffening and reduces the wall tension below the MGPI value.  Crucially, the correction is
proportional to $b^{2}$ and is therefore \emph{always} negative, irrespective of the sign of $b$ (equivalently, irrespective of whether $r_{e}$ is greater or less than $2a_{s}$), so the finite-range effect invariably reduces the wall tension relative to the MGPI prediction.

The wall tension at the MGPII level may exceed or fall below the mean-field value $\gamma_{0}$, depending on the relative magnitudes of the LHY and finite-range contributions. The crossover occurs when $8.511\,\alpha_{s}^{1/2} = 0.0063\,b^{2}$, at which point the two beyond-mean-field contributions mutually cancel and $\gamma_{\rm II}=\gamma_{0}$.  In the dilute limit where $\alpha_{s}\ll 1$ and $b\ll\alpha_{s}^{1/4}$, the typical hierarchy 
\begin{eqnarray}
\gamma_{0} < \gamma_{\rm II} < \gamma_{\rm I}
\end{eqnarray}
holds, consistent with the perturbative nature of both corrections~\cite{fu2003beyond}.  As the gas parameter is increased toward unity, for instance, by tuning the scattering length via a Feshbach resonance~\cite{Inouye1998,Cornish2000}, the beyond-mean-field contributions to the wall tension grow in magnitude.  In this regime, precision measurements of $\gamma$ could in principle provide
experimental access to both the LHY correction and the finite-range physics encoded in $b$~\cite{Blume2001}.

\section{Conclusions}
\label{sec4}

We have investigated the effects of QFs and finite-range interatomic interactions on the wall tension of a BEC confined by a hard wall at zero temperature. Starting from an energy functional that incorporates the LHY correction and a finite-range coupling constant, which lead to the modifications of the GP equation, namely, MGPI and MGPII, and obtained analytical expressions for the condensate wave function and the wall tension at each level.

The main findings of this work are as follows. (i) QFs, entering through the LHY term, stiffen the
condensate and narrow the density profile near the hard wall.  As a consequence, the wall tension is enhanced above its mean-field value by an amount proportional to $\alpha_{s}^{1/2}$, yielding
$\gamma_{\rm I}>\gamma_{0}$ unconditionally in the dilute limit.  At the same time, the healing length is reduced, $\xi_{\rm I}<\xi$, in agreement with the observed narrowing of the density notch.
(ii) Finite-range interactions, parametrized by the dimensionless quantity $b$, introduce a further correction to the wall tension that is proportional to $-b^{2}$ and therefore always negative.  The
finite-range effect thus partially offsets the LHY enhancement, resulting in the ordering $\gamma_{0}<\gamma_{\rm II}<\gamma_{\rm I}$ in the typical dilute regime. (iii) The two beyond-mean-field contributions mutually cancel when $8.511\,\alpha_{s}^{1/2}=0.0063\,b^{2}$, indicating that the wall tension can be tuned back to its mean-field value by choosing an
appropriate interatomic potential.  This cancellation condition provides a sensitive probe of the microscopic interaction.

These results demonstrate that the wall tension of a BEC at zero temperature is a sensitive and experimentally accessible observable for detecting beyond-mean-field physics.  As the gas parameter is
increased toward unity via Feshbach resonance tuning, the LHY and finite-range contributions become significant, and precision measurements of the wall tension could simultaneously constrain both
the LHY coefficient and the effective range of the interatomic potential.  Future extensions of the present framework include multicomponent BECs, where interfacial tension and immiscibility 
transitions are relevant, as well as finite-temperature effects and strongly correlated regimes such as self-bound quantum droplets in Bose mixtures.

\section*{Conflict of interest}
The authors declare no conflict of interest.

\end{document}